\documentclass[11pt]{article}

\usepackage{amssymb, amsmath, amsthm, mathrsfs}
\usepackage{graphicx, array, wrapfig, multirow, fullpage, caption, subcaption}
\usepackage{hyperref}

\begin{document}

\title{Revisiting the thermal and superthermal two-class \\ distribution of incomes: a critical perspective\thanks{Forthcoming in EPJ B}}
\date{\today}
\author{Markus P.\ A.\ Schneider\thanks{\href{mailto:markus.schneider@du.edu}{markus.schneider@du.edu}} \\ University of Denver}

\maketitle

\abstract{
This paper offers a two-pronged critique of the empirical investigation of the income distribution performed by physicists over the past decade.  Their finding rely on the graphical analysis of the observed distribution of normalized incomes.  Two central observations lead to the conclusion that the majority of incomes are exponentially distributed, but neither each individual piece of evidence nor their concurrent observation robustly proves that the thermal and superthermal mixture fits the observed distribution of incomes better than reasonable alternatives.  A formal analysis using popular measures of fit shows that while an exponential distribution with a power-law tail provides a better fit of the IRS income data than the log-normal distribution (often assumed by economists), the thermal and superthermal mixture's fit can be improved upon further by adding a log-normal component.  The economic implications of the thermal and superthermal distribution of incomes, and the expanded mixture are explored in the paper. \\

} 

\clearpage

\section{Introduction} \label{sect:intro} Since the publication of \cite{YakoDrag01-2} and \cite{YakoDrag01}, there has been persistent research effort by physicists into the apparent features of the income distribution.  The seminal study \cite{YakoSilva05} found that the observed income distribution -- based on individually-filed tax returns -- showed evidence of a two-class structure, in which the lower majority of the distribution appears to resemble the Boltzmann-Gibbs distribution.  Since this is the canonical result of classical thermodynamics, they dubbed this the {\it thermal} distribution of incomes.  The upper tail of the income distribution appears to follow a power-law, which \cite{YakoSilva05} named the {\it superthermal} portion of the income distribution. In the econophysics literature, these basic results have become the received wisdom upon which much subsequent research has been based (see \cite{Chatt07}, \cite{YakoRev07}, and \cite{YakoRoss09} for summaries), although the empirics have received relatively little critical review - until now.

For their part, economists have been generally skeptical of the evidence for a thermal distribution of incomes or the discrete two-class structure described by \cite{YakoSilva05}.  The finding of a power-law upper tail, on the other hand, generally resonates with economists' work (see \cite{Simon57} for example).  At least some of the disagreement between the relevance of the empirical findings by \cite{YakoSilva05} has to do with varying degrees of appreciation for statistical mechanics by economists.  As the short summary piece \cite{Cho14} illustrates, some economists are skeptical of the empirical findings but appreciate the meaning and power of statistical equilibrium, while others cannot let go of the idea that individual agency and attributes somehow make conclusions about the shape of the observed distribution of income impossible.  Part of the problem is that economists spent decades searching for a satisfying fit of a known statistical distribution with inconclusive results (see \cite{McD84}).\footnote{More correctly, the best fitting candidates where invariably too complicated to yield clear insight about the generating mechanisms. \cite{Dag77} offers something of an exception.}  However, finite mixture models received little attention even though there should have been ample evidence that they might be relevant.  Despite offering a critical perspective of the specific empirical results that physicists have honed in on, I am deeply appreciate of the approach physicists have taken and the general discovery of robust evidence for a multi-class structure of the observed income distribution.

Concurrent with -- and inspired by -- these empirical discoveries was the development of a statistical mechanics approach to money starting with \cite{YakoDrag00}.  The basic premise of this work was that the exponential distribution of the majority of incomes suggested a conservation law akin to the conservation of energy in thermodynamics.  Physicists quickly posited that the relevant conservation law must be the conservation of money.  Support for this idea was shored up using various agent-based models that showed that random exchanges of money between a fixed number of agents, during which money is neither destroyed or created in the act of exchange, result in a stationary exponential distribution of money (see \cite{YakoRev07} and \cite{YakoRoss09} for summaries of these modeling efforts).

As \cite{Durlauf05} saliently points out with respect to the hunt for evidence of complexity, showing that a simple model produces a feature which appears to be consistent with the observed data does not mean that the model correctly models real economic processes, unless it can be shown that it is the only model to produce this feature.  The fact that econophysicists have produced numerous agent-based simulations that result in a stationary exponential distribution has failed to convince many economists that the conservation of money -- or any more general conservation law -- is relevant to their field.  Critical responses to the econophysicists' research were provided by \cite{Lux06} and a scathing rebuttal by \cite{McCauley06}, a fellow physicist who expressed particular doubts about the relevance of conservation laws in economics.

If the goal is to find a model that is as simple as possible and no simpler, then the resounding response by economists has been that the proposed models are too simple.  But as indicated in \cite{Cho14}, economists are split among those that are conceptually supportive of the approach taken by econophysicists (like myself) and those that believe that attempting to uncover revealing features in distributional economic data is futile.  One of the issues is that compared to data in physics, economic data is noisy and sparse.  A salient complaint brought forth in \cite{Durlauf05} is that the general aversion to formal statistics in the physics community seem to have lead to conclusions about observed features in the data that on closer examination are not robust.  It is with respect to this last point that the present paper makes a contribution.

Assuming for the moment that the income data is robustly exponentially distributed, it is a fair to ask what this implies.  The key features that lead to the Boltzmann-Gibbs distribution as the stationary outcome of statistical mechanical system are that 1) there is a lower bound to the permissible observations, and 2) that the mean is constant.\footnote{The core narrative of classical thermodynamics was criticized along these lines by \cite{Jaynes57}.  Where it may not be a relevant critique of thermodynamics, it is an important point to modeling social systems.}  In classical thermodynamics, the first condition is trivially satisfied because negative energies are not defined.  Similarly, negative incomes are not economically meaningful, so this lower-bound constraint is not objectionable.  

The conservation of energy is a direct consequence of 2) because the size of the system is typically assumed to be fixed.\footnote{Permissible generalizations hinge on the system changing size relatively slowly compared to the rate at which equilibrium is reached, so that system size can be assumed de facto fixed during the relevant time interval.}  Without the assumption of fixed system size, there may be many other micro-dynamic interactions that are consistent with premises 1) and 2) -- and therefore would lead to a stationary exponential distribution -- but that do not necessarily imply the conservation of money or any other extensive property per se. 

Economists have long either explicitly or implicitly posited that the distribution of wages is close to log-normal based on the idea that the individual contributions to an employee's productivity (like experience, education, innate qualities, etc.) have a multiplicative effect, and that employees are remunerated according to their marginal productivity.  This idea is often credited to Gibrat (see \cite{Sutton97}), although \cite{Kalecki45} and others have shown that multiplicative processes can also lead to power-laws consistent with earlier speculations by \cite{Pareto}.  The important point brought up by the econophysicists' research is that processes that lead to a Gibrat distribution and processes that lead to a Boltzmann-Gibbs distribution are different in fundamental ways.  The apparent better fit of the exponential to the lower majority of the income distribution thus threatens a lot of economic theory, and for this reason alone its robustness should be explored seriously -- by economists, as well as physicists.

To be fair to economists, \cite{Lydall59} (among others, see \cite{McD84}) pointed out that incomes did not appear to be log-normally distributed, and some used this fact to argue against the distribution of wages reflecting differences in marginal productivities.  This empirical insight has been largely ignored by most economists in the decades since and most never considered explicitly working out the distributional implications of their models of wage determination.  A relevant exception is the formal application of statistical mechanics to markets provided by \cite{Foley94} and extended to a simplified labor market in \cite{Foley96}.  The latter appears to offer the only formal explanation for an exponential distribution of wages in an economic model; but by the author's own admission, it only does so under extremely unrealistic assumptions.  A promising recent contribution by economists is \cite{ShaikhEtAl14}, which brings more grounded economic reasoning to the same empirical analysis relied upon in the physics literature.

The findings presented by econophysicists further suggest that the overall distribution of incomes provides evidence for a two-class structure, as made explicit by the notion of a thermal / superthermal distribution (\cite{YakoSilva05}).  Again, due attention to what has been done by economists to explain similar phenomena has been missing from the econophysics literature.  The insight that the labor market is segmented was formalized by \cite{Weitz89}, though little attention was paid to the distributional implications the characteristics of each segment suggest.  Recent theories about the role of search frictions and matching in the labor market have also produced models that allow for multiple equilibria (see \cite{Rogerson05} for a summary), which may be consistent with the multi-class structure in the observed distribution.\footnote{Economic models that are compatible with multiple equilibria are rarely interpreted as leading to the coexistence of different equilibria in a heterogeneous population, but they may at least address the issue that different segments of the population experience fundamentally different mechanisms leading to equilibrium.}  While no economic model directly predicts econophysicists' findings, it is a mistake not to explore to what extent a thermal / superthermal structure of the income distribution can be reconciled with these existing theories.

The point of the present paper is not, however, to try to outline the lacuna between economic theory and econophysicists' empirical findings, but to focus more narrowly on the robustness of the empirics themselves.  This is a necessary (but not sufficient) step towards understanding labor market outcomes and reconciling physicists' and economists' parallel investigations.  Econophysicists' fit of the exponential distribution is based on a graphical analysis and rules of thumb of how data from a particular distribution behaves.  While this type of analysis is insightful as a first step, it should not be relied upon as the last word in fitting a distribution to the data.  Furthermore, these techniques work extremely well when the researcher is dealing with a large amounts of noiseless data, which income data is not.  It is worth noting that work like \cite{McD84} and \cite{Wu03} presented sophisticated analyses (the latter based on algorithmically fitting a general exponential maximum entropy density to an observed income distribution) that by implication ruled out the appropriateness of the exponential distribution in favor of much more complex candidates.  Although it should be noted that neither \cite{McD84} and \cite{Wu03} considered finite mixtures as proposed by physicists.

This paper begins with critical remarks regarding the techniques used to identify the thermal and superthermal distribution of incomes.  On purely analytical grounds it is possible to show that the simple graphs of the complementary cumulative distribution of the data may not allow the researcher to distinguish between exponentially and log-normally distributed data.  Second, the results of a cursory statistical analysis using popular measures of fit is presented, which casts further doubt on the fit of the exponential portion of the thermal and superthermal distributional model proposed by \cite{YakoSilva05}.  Specifically, the results presented in this paper suggest that a third class should be added to the distributional model.  Thus, the final results are not a vindication of conventional economic thinking either, but rather prompt for both a more in-depth empirical analysis as well as the need to answer the theoretical questions that these findings bring to light.

\section{Weakness of the Graphical Argument} \label{sect:crit} At the heart of \cite{YakoSilva05}'s analysis is a graph of the empirical complementary cumulative distribution ({\it ccdf}) on a plot with linear-log axes.  The complementary cumulative distribution function ({\it ccdf}, given by $1 - F_{X}[x]$ where $F_{X}[x]$ is the cumulative distribution function or {\it cdf})\footnote{Following common convention -- although breaking with \cite{YakoRev07} -- lower-case functions refer to the probability density function ({\it pdf}) and capital letter functions refer to the cumulative distribution function ({\it cdf}) of a random variable.  E.g., if $X$ is a random variable supported on $x\in(0,\infty)$, then $F_{X}[x] = \int_{0}^{x} f_{X}[\xi] \,d\xi$ is its {\it cdf}.} of the Boltzmann-Gibbs distribution would plot as a straight line on this kind of graph, with the slope given by $-\beta$.

Instead of plotting the {\it ccdf} against nominal income values, \cite{YakoSilva05} normalized the reported incomes by dividing them by the temperature of the exponential component of the observed distribution.  Under the assumption that the observed data is exponentially distributed, the normalization using the temperature amounts to dividing the observed incomes by their sample mean.  Since \cite{YakoSilva05} posit a power-law tail for high incomes, they had to correct for the mean income reflecting this and hence they use the maximum likelihood (ML) estimate of the temperature instead.  This normalization using the temperature allows the comparison of income distributions from different years on the same graph.  The fact that the normalization procedure appears to collapse the {\it ccdf}s from different years onto the same line is the key finding that \cite{YakoSilva05} use to support their claims of exponentiality: collapse suggests only scale changes from year to year, while linearity implies exponentiality.

\begin{figure}
\centering
\begin{subfigure}{0.45\textwidth}
\includegraphics[width=\textwidth]{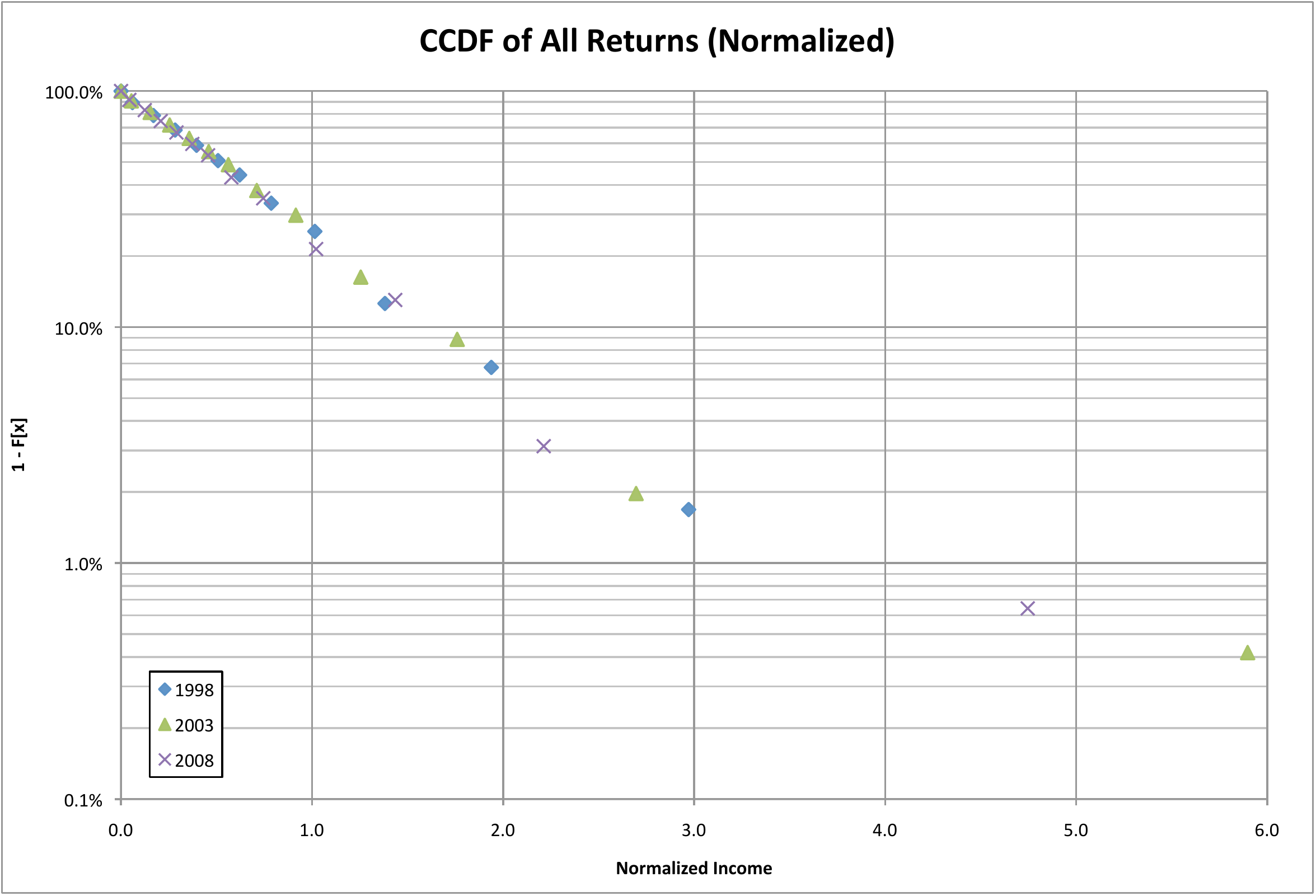}
\end{subfigure}
\begin{subfigure}{0.45\textwidth}
\includegraphics[width=\textwidth]{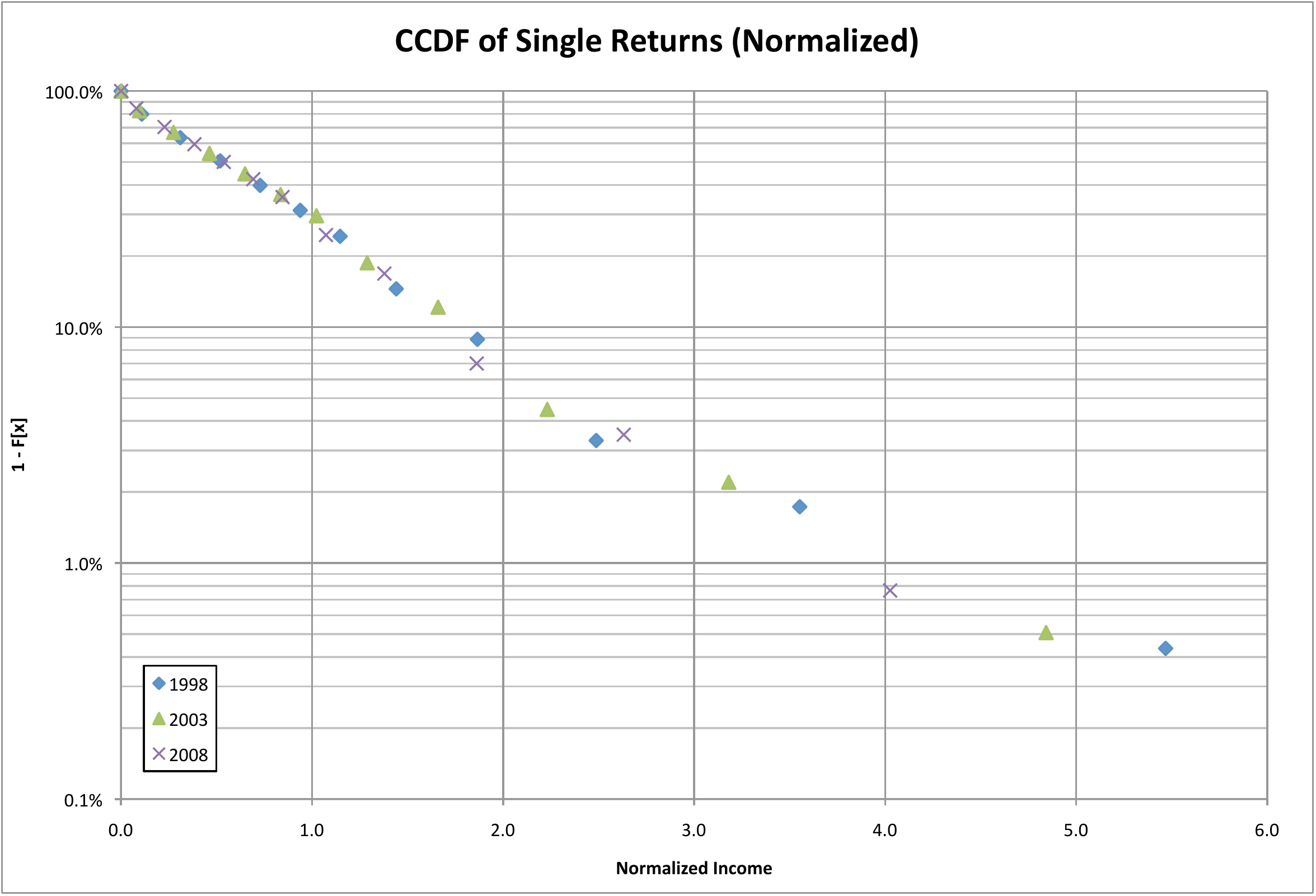}
\end{subfigure}
\caption{The complementary cumulative distributions ({\it ccdf}s) over normalized income values for 1998, 2003, and 2008 based on publicly available IRS reports of number of returns and Adjusted Gross Income (AGI).} \label{fig:CCDFs}
\end{figure}

The graphs in figure \ref{fig:CCDFs} are created exactly according to this procedure and clearly show that the income distributions from 1998, 2003, and 2008 collapse onto the same {\it ccdf} when incomes are normalized.  The graph on the left shows the income distribution for all US tax returns, while the one on the right is based on individually-filed returns only.

Note that both graphs in figure \ref{fig:CCDFs} appear to show an apparently linear relationship between {\it ccdf} and normalized income value up to a normalized income value of 3.0 or 4.0.  This linearity is the core piece of evidence that physicists cite to suggest that the lower portion of the income distribution might be exponential.  It is not clear how trustworthy the visual appearance of linearity really is, however.  

\cite{YakoRev07} further implicitly suggests that log-normally distributed data would be unlikely to collapse onto the same {\it ccdf} after normalization unless one shows that the shape parameter ($\sigma$) was relatively constant from year to year.\footnote{Given $X \sim \textrm{lgN}[\mu,\sigma]$, the distribution of $Y = X / \left<X\right>$ is a log-normal with a mean of one whose shape depends only on $\sigma$.}  (Similarly, normally distributed data would only collapse after normalization by its mean if the coefficient of variation was constant across years.)  The estimates presented in this paper suggest that if a log-normal component is included in the mixture model fit to the data, $\sigma$ for that component does not vary much from 1998 to 2008 (see table \ref{tbl:ParamEst} in the appendix).  Only the close adherence to a straight line therefore distinguishes exponential from log-normally distributed data, and that assessment may be more difficult to make visually using economic data than physicists are willing to admit.

In the mid-range of the support, both log-normal and normal distributions can appear close to linear on a log-linear graph.\footnote{The log-linear plots in \cite{ShaikhEtAl14} quite clearly show a soft but discernible concave curvature above the straight line of the {\it ccdf} for low incomes and mild convex curvature are mid-range incomes.}  If there are sufficient distortions at the low-end of the support of the observed distribution (perhaps due to noise or disequilibrium effects) or due a missing mixture component that is particularly important at low incomes, then visual inspection may lead to incorrect inferences.  Visually distinguishing between exponentially and log-normally distributed data especially is only reliable when the observed data is noiseless and plentiful; requirements that the income data may not meet as satisfactorily as data generated by a well-designed experiment in physics.

A further point is that \cite{YakoSilva05} assumed a hard boundary between exponentially and power-law distributed incomes.  There is no economic rational for assuming that all income above a specific single value strictly follow a different distribution than those below it.  More recent work like \cite{BanerjYako10} has addressed this issue by positing asymptotically exponential and power-law behavior towards the extremes of the support.  Specifically, \cite{BanerjYako10} (and similarly \cite{FiaschiMars12}) allow additive and multiplicative diffusion processes to overlap so that the overall stationary distribution interpolates between the two processes' signatures.  Unpublished work by \cite{BloomquistEtAl} took up this modeling consideration in a different way by allowing the respective distribution simply to overlap.  The same approach is taken in this paper because I posit that the important insight from the econophysics literature is the evidence for finite mixture components, each of which represents a distinct generating mechanism.

According to \cite{BloomquistEtAl}'s estimates, the lower bound to the power-law distribution was like to be \$44,000, considerably below the apparent transition in the log-linear plot of the {\it ccdf}.  The mixture they used is of the form (\ref{eq:mixexppwr}), which allows the distributional components to overlap.  Intuitively, this form of the mixture allows any observation of $x \geq k$ to come from either the exponential contribution to the mixture or the power-law tail, but not both.\footnote{Since the power-law component is supported on $[k,\infty]$ with $k > 0$, all observation of $x < k$ must come from the exponential component in the mixture.}

\begin{equation} \label{eq:mixexppwr} f_{\textrm{mix}} \left[x\right] = A \, f_{\textrm{exp}} \left[x \right] + (1-A) \, f_{\textrm{pwr}} \left[x \right]
\end{equation}

\noindent where $f_{\textrm{exp}} \left[x \right]$ refers to the {\it pdf} of an exponential distribution with parameter $\beta$ and $f_{\textrm{pwr}} \left[x \right]$ refers to the {\it pdf} of a power-law distribution with lower bound $k$ and shape parameter $\alpha$.  The parameter $A \in [0,1]$ provides a direct estimate of the contribution of the exponential component to the pooled distribution in \%-terms.

\section{Testing Distributional Fit} \label{sect:fit}  Having argued that the visual methods used by econophysicists may not be as robust is desirable, an alternative analysis is necessary.  A still quite cursory analysis of distributional fit is presented in this section.  A much more rigorous formal analysis along these lines should be conducted using the uncensored tax data (not publicly available) spanning as many years as possible.  However, to provide a preliminary verification or refutation of econophysicists' main findings, two alternative distributional models are fit to the IRS income data for three different years.  The fit of each distributional model is assessed using common fit criteria as well as the index of informational distinguishability generalized in \cite{Soofi95}.\footnote{A caution against the dismissal of formal assessments of fit: \cite{Soofi95}'s methodology rests on information theoretic foundations and can easily be reconciled with the work of E.\ T.\ Jaynes, whose approach I would describe as a physicist's antidote to formal statistics.}

In accordance with \cite{Soofi95}, the parameters for each distributional model were estimated by maximizing the likelihood.  Each distributional model specified by the ML parameter estimates was then used to calculate the informational distinguishability index ($ID$), calculated according to (\ref{eq:ID}), relating it to the observed distribution of income.  This index takes a value between 0 and 1, with 0 indicating that two distributions are informationally indistinguishable from each other.  Comparing the distributional models to the observed distribution of income therefore means searching for the model with the lowest $ID$ score.  The incremental contribution of one model compared to another model that imposes less on the information provided by the data can be assessed using the relative information, $RI$, calculated using (\ref{eq:RI}).  $RI$ also ranges from 0 to 1, with 0 indicating that the change in model did not improve in the use of the information provided by the data.\footnote{Specifically that if both models are constructed as solutions to a maximum entropy program, the additional constraints implied by the more complex model are not binding.}

\begin{equation} \label{eq:ID} ID\left[p:p^*\right] = 1 - \exp\left[D_{KL}[p:p^*]\right]  \end{equation}

\noindent where $D_{KL}$ is the Kullback-Leibler divergence between the observed distribution $p$ and the distributional model $p^*$.  In (\ref{eq:RI}), $p^*_1$ represents the model that asks less of the data.  Fitting the model represented $p^*_2$, which asks more of the data (either in terms of more parameters that need to be estimated on in how restrictive the modeling features are, or both), implies $ID\left[p:p^*_2\right] \leq ID\left[p:p^*_1\right]$.  If $ID\left[p:p^*_2\right] = ID\left[p:p^*_1\right]$, then the fit of $p^*_1$ and $p^*_2$ are informationally indistinguishable and $RI = 0$.
 
\begin{equation} \label{eq:RI} RI = 1 - \frac{ID\left[p:p^*_2\right]}{ID\left[p:p^*_1\right]}  \end{equation}

The Kolmogorov-Smirnov statistic ($D_{KS}$), which is based on the difference between the observed cumulative distribution and the {\it cdf} implied by the distributional model, was also calculated to assess fit.  Finally, the Aikake and Bayesian Information Criteria ($AIC$ and $BIC$ respectively) are cited since they are trivial to calculate from the value of the maximized likelihood function and are standard measures for evaluating model fit, especially when parsimony is a serious consideration.  A detailed description of $D_{KL}$, $D_{KS}$, and the information criteria can be found in the appendix.

\subsection{Data} \label{sect:data} To make the present work comparable to \cite{YakoSilva05} and other research by physicists, the same publicly available tax returns data provided by the Internal Revenue Service (IRS) is used.\footnote{Estimates for returns across income brackets by filing status are available publicly via the Statistics of Income (SOI) research division at the IRS ({\tt http://www.irs.gov/uac/Tax-Stats-2}).}  It now appears to be common practice by econophysicists to look only at the distribution of single tax payers under the belief that this avoids interdependencies caused by household decision making.  From an economists standpoint, it would be desirable that either the same distributional model fits both the observed distribution of single tax payers' incomes and all incomes, or that the distribution based on all returns can be easily explained in terms of the simpler distribution attributed to single returns.

For example, if the distribution of individually-filed returns is largely exponential, the pooled distribution of all returns should have a substantial exponential component plus a component that reflects individual income earners joining into a households reporting income that is in part influenced by household decision making.  \cite{YakoSilva05} actually suggested that the distribution of household incomes can be satisfactorily modeled as the convolution of independent exponential incomes (i.e., as a $\Gamma[2,\beta]$ distribution).  If this were indeed the case, then there should be no reason to separate the distribution of incomes reported by single tax payers from those filing jointly, or as divorcees and widows/ers, etc.  In an effort to shed some light on how much of a difference using only a subset of returns makes, results for both the income distribution of individual tax payers as well as for the distribution of all reported incomes will be presented.

\subsection{Alternative Mixtures} \label{sect:altmix} In total three distinct mixture models will be fit to the observed distribution of incomes.  The first is the one based on \cite{YakoSilva05} that combines an exponential component with a power-law tail as given in (\ref{eq:mixexppwr}).  In this model, the parameter $A \in [0,1]$ is a direct estimate of the \% of the distribution attributed to the exponential.  This model is contrast to a similar mixture where the exponential component, $f_{\textrm{exp}}[x]$, is replace by a log-normal component, $f_{\textrm{lgN}}[x]$ with parameters $\mu$ and $\sigma$.  The parameter $A$ again appears in this mixture model and has qualitatively the same meaning, this time providing a direct estimate of what contribution the log-normal component makes.  The log-normal / power-law mixture is a nod to economists' common belief that much of the income distribution is log-normal while also incorporating that economists' have long suspected that high incomes may exhibit power-law behavior (see \cite{Pareto} and \cite{Simon57}).

It is quite plausible that this either / or approach to modeling the observed income distribution neglects the possibility that there is a third segment in the labor market, and that each segment has characteristics consistent with either one or another distribution in equilibrium.  For example, the dual labor market theory of segmentation as per \cite{GordonEtAl73} provides a narrative that could be consistent with a mixture that shows three distinct components.  Very recent models in econophysics, like \cite{Yuqing07}, also suggest that more complicated mixtures capturing transition dynamics are possible.  While no effort is made here to try to distinguish between the applicability of one explanation or the other, I simply investigate the plausibility of a third contributing component.

To do so, the distributional model given by (\ref{eq:mixexplgNpwr}) will also be fit to the data.  The third component in (\ref{eq:mixexplgNpwr}) is a log-normal that would indicate a labor market segment which allocates incomes consistent with standard economic theory, and from a statistical mechanics perspective would imply that the interaction process has other constraints than simply mean preservance.  The mixture given by (\ref{eq:mixexplgNpwr}) explicitly parameterizes the contribution to the overall mixture of the exponential component (through $A \in [0,1]$) and the power-law tail (through $B \in [0,1]$), and hence the complimentary contribution of the log-normal component as $(1 - A - B)$.

\begin{equation} \label{eq:mixexplgNpwr} f_{\textrm{mix}} \left[x \right] = A \, f_{\textrm{exp}} \left[x \right] + (1- A - B) \, f_{\textrm{lgN}} \left[x \right] + B \, f_{\textrm{pwr}} \left[x \right] \end{equation}

It is important to emphasize that the claim is not that (\ref{eq:mixexplgNpwr}) provides the best fit of the data in some absolute sense, but rather that the graphical analysis relied upon by econophysicists cannot conclusively rule out this more complex mixture.  Some of the fit criteria used in the analysis presented in this paper are specifically designed to balance added complexity with better fit.  If according to these fit criteria the Boltzmann-Gibbs / Gibrat / Pareto mixture fits the income distribution better than the Boltzmann-Gibbs / Pareto mixture, then it truly suggests that the latter was too simple a description of the data.\footnote{The Boltzmann-Gibbs / Gibrat / Pareto mixture is only one plausible candidate to replace the thermal / superthermal distribution suggested by \cite{YakoSilva05}.  While other combinations have been tested informally and the preliminary results suggest that it is the most reasonable candidate, the present study is far too cursory to conclusively show that it is the best candidate.}

\section{Results} \label{sect:res} The three distributional models described in the previous section are fit to the data provided by individually-filed returns (single) and all returns for all three years (1998, 2003, 2008) to assess their fit of the observed distribution.  A series of common fit criteria are calculated for each model specified using the maximum likelihood (ML) parameter estimates.  All the fit criteria discussed in section \ref{sect:fit} indicate the same basic result: the exponential / power-law combination (\ref{eq:mixexppwr}) provides a better fit of the IRS income data for both single and all returns than a combination of the log-normal distribution with a power-law tail.  While this appears to be vindication for thermal / superthermal distribution of incomes, the expanded mixture incorporating both an exponential and a log-normal component with a power-law tail (\ref{eq:mixexplgNpwr}) provides an even better fit by all measures.  The formal results therefore support the conclusion anticipated in the previous section that the thermal distribution alone is too simple a description of the lower majority of the income distribution, and that the added complexity of the Boltzmann-Gibbs / Gibrat / Pareto mixture is justified.

\begin{figure}
\centering
\includegraphics[width=0.9\textwidth]{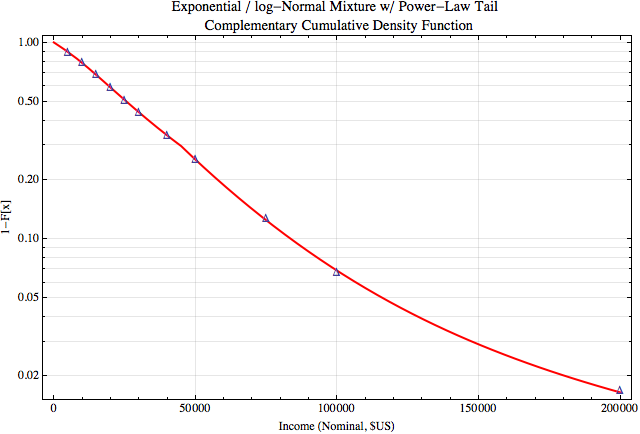}
\caption{The {\it ccdf} parameterized using the ML estimates for the 1998 distribution of AGI reported on all IRS returns.} \label{fig:CCDFmix}
\end{figure}

To illustrate the claim that the mixture given by (\ref{eq:mixexplgNpwr}) cannot be ruled out using a linear-log plot of the {\it ccdf}, figure \ref{fig:CCDFmix} shows that the {\it ccdf} of the exponential / log-normal / power-law mixture fitted to the income data may appear linear for incomes up to \$100k.  As suggested earlier, as long as the shape parameter $\sigma$ does not vary significantly across years, normalizing by the mean $x$ will collapse both components of (\ref{eq:mixexplgNpwr}) onto standardized curves if the weight of each contribution to the mixture also remains stable across years -- i.e., if $A$ is relatively constant.  Given the estimates produced for this study, the normalization suggested by \cite{YakoSilva05} does not rule out this kind of mixture.  Estimates for $\sigma$ range from $0.615$ for the 1998 singles data to $0.650$ for singles data in 2003, and estimates for $A$ from 70\% for singles data in 1998 to 79\% for 2003 (estimates in 2008 fell between these values).  These apparently sizable variations in the parameters are insufficient to disrupt the collapsing of the normalized distributions.  The more formal fit criteria confirm that (\ref{eq:mixexplgNpwr}) provides a better fit of the lower majority of the income distribution for both individually-filing tax payers as well as when fit to the pooled distribution of all tax payers than the exponential / power-law model given by (\ref{eq:mixexppwr}).

The relative information gain associated with going from the thermal / superthermal distribution to the exponential / log-normal / power-law mixture ranges from 0.66 for the 2008 data to 0.76 for the 1998 data (all returns).  For singles, the improvement in capturing relative information is somewhat lower - as might be expected - with $RI$ ranging from 0.36 to 0.53.  In either case, $RI$ indicates that the more complex mixture uses the information provided by the observed data much better.  Alternatively, the $AIC$ and $BIC$ are improved by 1.2 million or more for all respondents - and 150,000 for singles - suggesting that the added complexity of including both exponential and log-normal components in the mixture is more than justified by the improved fit.  This result would not be apparent from the distributional graphs used by \cite{YakoSilva05}, but as figure \ref{fig:CCDFmix} showed it is also not inconsistent with the key features discovered using the graphical analysis.

\begin{figure}
\centering
\begin{subfigure}{0.45\textwidth}
\includegraphics[width=\textwidth]{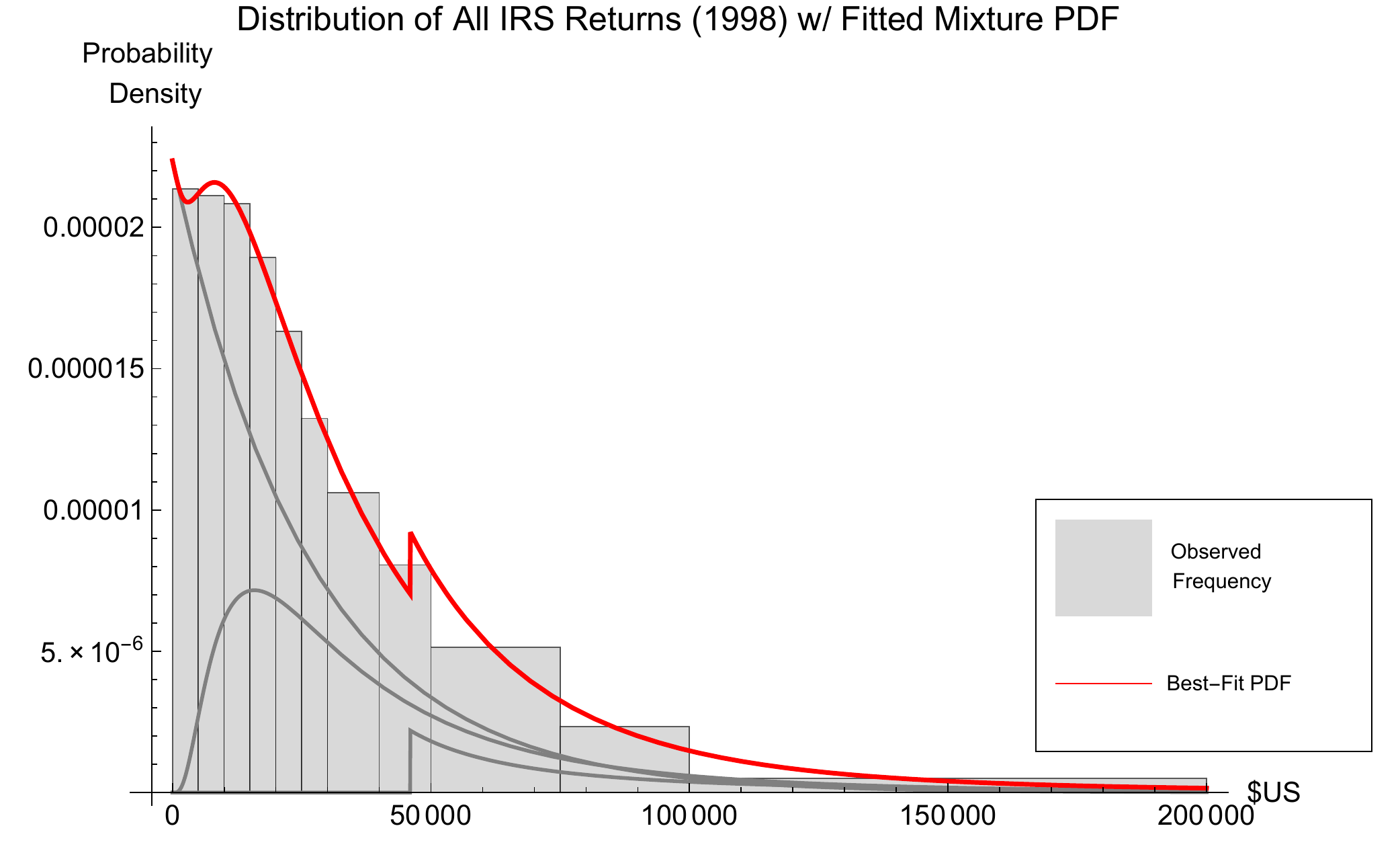}
\end{subfigure}
\begin{subfigure}{0.45\textwidth}
\includegraphics[width=\textwidth]{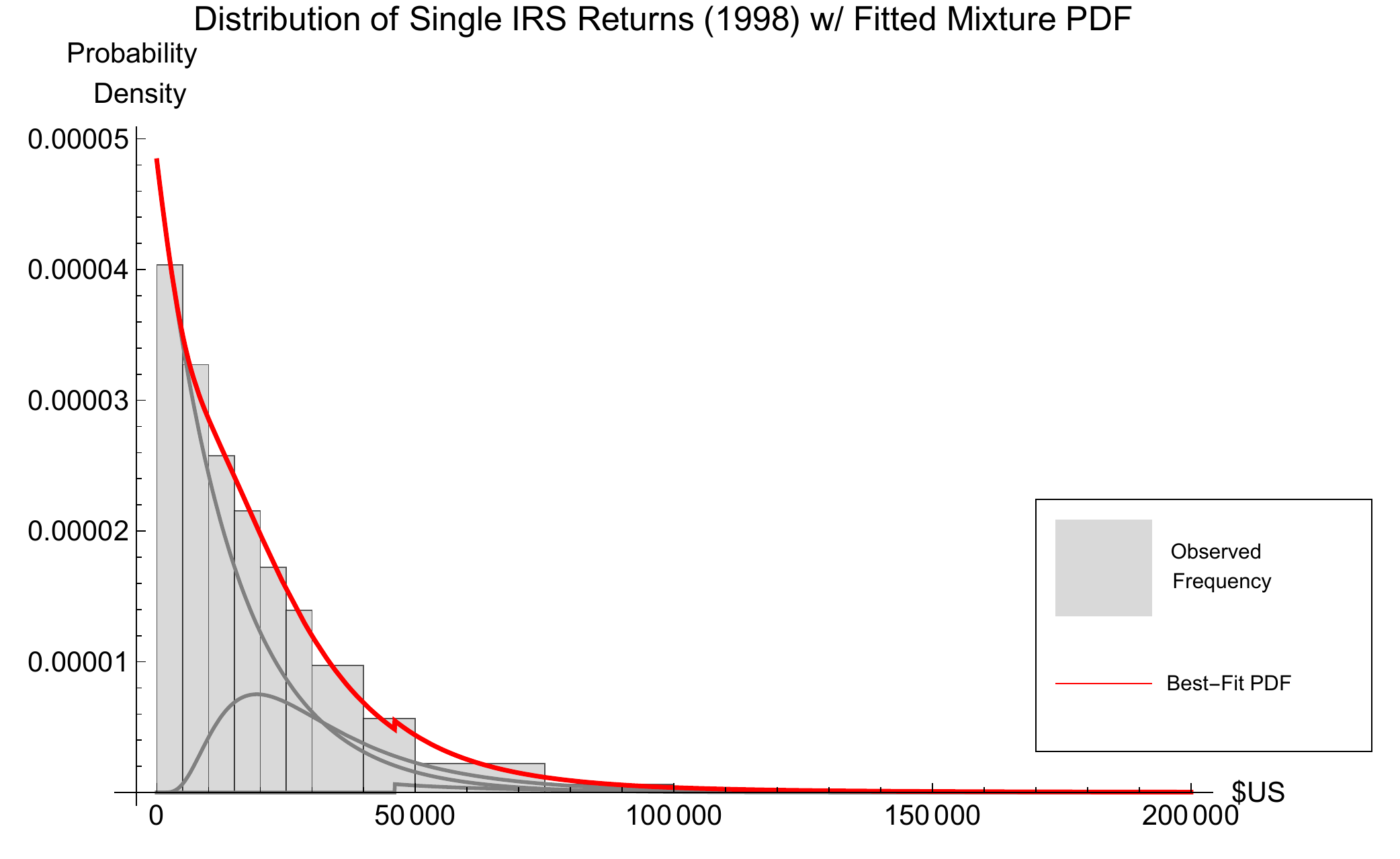}
\end{subfigure}
\caption{The fitted {\it pdf} of the thermal / log-normal / superthermal mixture fitted to the observed distributions of all (left) and single (right) IRS returns in 1998.  The scaled mixture components are shown separately in gray.} \label{fig:fitPDFs}
\end{figure}

The Kullback-Leibler divergence between the data and the fitted exponential / power-law distribution is 0.0078 or greater for all respondents and 0.0030 for singles.  By comparison, fitting an exponential / log-normal / power-law distribution mixture to the data reduces $D_{KL}$ to 0.0026 or less for all respondents and 0.0022 or less for single respondents.  The Kolmogorov-Smirnov statistic is improved by almost a factor of 10 when the more complex mixture is used.  The best fitting data for the thermal / superthermal mixture was 2003, when the $D_{KS}$ was 310.  The worst fitting year for the more complex thermal / log-normal / superthermal mixture was 2008 with a $D_{KS}$ of 46.7.  For the singles return data, the improvement in $D_{KS}$ is only slightly less dramatic with the best year for the exponential distribution achieves a value of 290 and the worst year of the exponential / log-normal / power-law mixture achieves a 55.8.\footnote{While these values show a huge improvements in fit, it should be noted that they still likely justify formal rejection of fit. However, establishing appropriate critical values is difficult and was not undertaken for this study since the $D_{KS}$ is treated as a secondary criteria.}  The ML parameter estimates for each model and all calculated fit criteria can be found in tables \ref{tbl:ParamEst} and \ref{tbl:FitCrit} (respectively) in the appendix.  To help illustrate what ``good fit'' may look like, figure \ref{fig:fitPDFs} shows the histogram of the income data for 1998 with the fitted {\it pdf} superimposed for both all incomes (left) and single tax payers' incomes (right).

The results suggest that the graphical analysis relied upon by \cite{YakoSilva05} has likely led to an oversimplification of the distributional model from which explanations for the observed distribution of incomes are being developed.  The expanded mixture fit to the income data in this study matches the features identified by \cite{YakoSilva05}: it also to has a {\it ccdf} that appears linear on a linear-log plot and collapses when the data is normalized (given that the parameter estimates of $\sigma$, $A$, and $B$ vary only modestly).

In fair terms, this means that the visual analysis relied upon by econphysicists may have misled them about the magnitude of the exponential contribution to the observed distribution.  The results presented in this paper support the finding that the exponential component contributes significantly to the observed distribution (59\% in 1998 up to 67\% in 2008 based on ML estimates of the exponential / log-normal / power-law model), but this contribution is much smaller than the $>$90\% proposed by \cite{YakoSilva05}.  The estimated contribution of the power-law tail, on the other hand, is of a consistent magnitude with previous findings (8\% in 1998 down to 3\% by 2008).  The remaining 30\% of incomes reported to the IRS appear to be consistent with a log-normal component and suggest they originate from a process that was prematurely ruled out by econophysicists.  Looking only at single returns, the ratios change, but the qualitative insight does not: roughly 20\% of the distribution of single returns is allocated to the log-normal component in the complex mixture.  

The different composition of the pooled distribution of the incomes of single tax payers versus that of all incomes may be related to the age and career-stage of the income earners included in each.  Across the three years examined, 63\% of all tax payers' incomes appear to be exponentially distributed, while that proportion increases to an average 75\% when only single tax payers are considered.  This seems to indicate that a larger proportion of single tax payers are subject to a labor market in which their individual attributes do not influence their pay.  Given that single tax payers are likely younger and may not have started their careers, it should be expected that more of them hold entry-level jobs that require little demonstrated skill and generally do not compensate employees according to their acquired human capital.  This underlines the major contribution of this line of research by econophysicists: the labor market does seem to be segmented in such a way that different segments are governed by fundamentally different operational rules.  Furthermore, different parts of the labor force participate in these segments in different proportions and are therefore likely to have very different experiences.

\section{Conclusion} \label{sect:conc} Econophysicists have made an insightful contribution to understanding the distribution of income.  This paper aims to refocus what their contribution is: it appears that the exponential plays a significant role in the description of the observed distribution -- and the onus is on economists to explain how it arises -- but the equally economically significant insight is that the observed distribution suggests that different generating mechanisms govern the distribution of income for different labor market segments.  Furthermore, the evidence presented here shows that the thermal / superthermal distribution may be an oversimplification and that a third log-normal contribution should be considered.  The fact that the observed aggregate distribution shows signs of multiple contributing components -- each of which originates from a fundamentally different generating mechanism -- is a major finding that is reinforced by the results presented in this paper.

Given the results, economists are not off the hook to explain why a sizable portion of the pooled distribution of incomes appears to behave like a Boltzmann-Gibbs distribution.  More generally, economists should revisit theories of segmentation, be it by considering generalized modeling in econophysics like \cite{Yuqing07} or exploring the statistical mechanics of  models that incorporate search frictions and allow for multiple equilibria (see\cite{Rogerson05} for a summary).  It seems likely that we might find that in heterogeneous populations, there are labor market processes that lead to the coexistence of a finite number of coexistent statistical equilibria.

Conversely, econophysicists need to continue exploring generating mechanisms that can explain the observed segmented distribution, including a significant log-normal component.  They should, however, abandon the notion of a conservation law governing the allocation of money incomes, since it is an even harder to sustain idea for only one labor market segment, and does not seem strictly necessary.  Re-focusing their modeling efforts on mean-preserving dynamics may prove more fruitful and resonate better with economists.  A good starting point for this work would be to revisit theories of labor market segmentation (perhaps starting with the formal model by \cite{Weitz89}) and explicitly deriving the statistical mechanics of such a model as done in \cite{Foley96} for a simplified labor market.

A more serious empirical investigation is also warranted.  As the statistical rejection of fit indicates, the model presented in this paper is not the final word on how best to describe the observed distribution, although it is a stride in the right direction.  Prematurely jumping from these results to modeling exercises may yield unjustified conclusions as may have already occurred (also suggests by \cite{Durlauf05} with respect to complexity research).  To address the skeptics of formal statistical analysis: despite the known weaknesses, avoiding formal analysis does not insulate the researcher from relying on similar or even weaker methods if one does not take care to consider the problematic features of economic data.  The time seems ripe to encourage more critical collaborations between economists and physicists along both empirical and theoretical tracks -- in no small part in order to critically revise what we think we know from the early literature.


\bibliographystyle{plain}
\bibliography{boltzmann.bib}
\clearpage

\section*{Appendices} \label{sect:app}

\subsection*{Fit Criteria} For completeness, the fit criteria central to this investigation will now be briefly discussed, although the reader is likely familiar with these.  The Kullback-Leibler divergence measure is a distance measure that quantifies the additional message length required to describe the actual distribution generated by $p$ using the ``incorrect'' distribution function $p^*$.  The Kullback-Leibler distance, $D_{KL}$, is given by (\ref{eq:KL}), where $S[p,p^*]$ is the cross-entropy with probability density $p$ induced by the observed data and $p^*$ representing the {\it pdf} of a candidate distributional model.

\begin{equation} \label{eq:KL} D_{KL} = S[p,p^*] - \tilde{S} \end{equation}

$$ S[p,p^*] = \sum^{k}_{i=1} \frac{f_i}{n} \; \ln\left[\frac{P^*_i}{\Delta x_{i}} \right] $$

\noindent where $f_i$ is the number of observations in the $i^{th}$ bin and $P^*_i$ is the probability of an observation occurring in the $i^{th}$ bin under the assumption of the particular distributional model being considered.  This formulation assumes that there are $k$ bins of width $\Delta x_{i}$.  The probability $P^*_i$ is evaluated using the {\it cdf}, $F^*[x \, | \, \boldsymbol{\theta}]$, consistent with the reference density $p^*$ ($b_i$ is the $i^{th}$ bin's lower bound).

$$ P^*_i = F^*[b_i + \Delta x_{i} \, | \, \boldsymbol{\theta}] - F^*[b_i \, | \, \boldsymbol{\theta}] $$

Another common fit criteria is the Kolmogorov-Smirnoff (K-S) statistics, which compares the cumulative distribution of the data to the {\it cdf} implied by the model.  The largest absolute divergence between the two is used to assess fit, (\ref{eq:KS}).  A major advantage of the K-S test is that it does not rely on binned data.

\begin{equation} \label{eq:KS} D_{KS} = sup \Vert \; F[x_i] - F^*[x_i \, | \, \boldsymbol{\theta}] \; \Vert \end{equation}

\noindent where $F[x_i]$ is the relative cumulative frequency of the observation $x_i$ and $F^*[x_i \,|\, \boldsymbol{\theta}]$ the {\it cdf} based on the model being tested evaluated at $x_i$.  The sampling distribution of the Kolmogorov-Smirnoff measure is $sup \Vert \, B \left[F^*[t]\right] \, \Vert$, where $B[t]$ is the Beta distribution.  At a significance level of 5\%, the critical value is $1.358$ meaning that if $\sqrt{n}D_{KS}$ is greater than 1.358, the data suggests there is less than a 5\% probability of a Type I error if ``$H_0: \textrm{Fit}$'' is rejected.

Information criteria are another approach often used to test model specification and particularly over-parameterization.  If a parsimonious description of the data is being sought, then a criteria must be introduced that directly penalizes models for increasing the number of parameters that need to be estimated.  The Bayesian and Akaike Information Criteria (BIC and AIC respectively) are standard measures for model selection that are designed to select the most parsimonious model.

\begin{equation} \label{eq:BIC} BIC = -2 \, \ln L + \kappa \, \ln[n] \end{equation}

\begin{equation} \label{eq:AIC} AIC = -2 \, \ln L - 2 \, \kappa \end{equation}

The first term of these information criteria is determined by the value of the log-likelihood, $\ln L$, evaluated at the ML parameter estimates.  The second term penalizes the criterion for model complexity as captured by the number of parameters $\kappa$ (AIC), or model complexity and amount of available data (BIC).  In either case, a smaller BIC or AIC is always desirable.  A model that decreases the value of the BIC compared to another model by 10 or more is considered to provide a notable improvement, suggesting that whatever additional complexity was imposed is justified by the improvement in fit.

\subsection*{Likelihood Estimation} The IRS data used in this study comes from the tables published on irs.gov.  For each category of reported adjusted gross income (AGI), a total of reported income minus deficits is given in addition to the total number of returns falling into the category (e.g.\ AGI reported as between \$1 and \$5,000).  Using these two numbers, the average income for each category was calculated together with the proportion of returns falling into the category.  The average and the weight together allow for a calculation of the mean income, and the weight itself is taken as the best estimate of the probability mass associated with each category.  Only returns listing an AGI of \$1 or greater were considered.

\par The likelihood was constructed using the probability of making an AGI observation in the interval $[a_k, a_k + \Delta_k)$ (where $a_k$ is the $k^{\textrm{th}}$ bin's lower bound and $\Delta_k$ is the bin-width), (\ref{eq:prob}), assigned by the distribution whose {\it cdf} is given by $F[x; \boldsymbol{\theta}]$.  If the frequency of observations falling into this interval is $\varphi_k$, then the log-likelihood is given by (\ref{eq:lglik}).

\begin{equation} \label{eq:prob} P\left[ x \in [a_k, a_k + \Delta_k) \right] = F[a_k + \Delta_k; \boldsymbol{\theta}] - F[a_k; \boldsymbol{\theta}]
\end{equation}

\begin{equation} \label{eq:lglik} lgL = \sum^{m}_{k = 1} \varphi_k \, \ln\left[ F[a_k + \Delta_k; \boldsymbol{\theta}] - F[a_k; \boldsymbol{\theta}] \right]
\end{equation}

\noindent where $m$ is the number of bins.  The ML estimators, $\boldsymbol{\tilde{\theta}}$, are chosen so that they maximize $lgL$, which was accomplished in the present study using a built-in algorithm in {\it Mathematica} called {\bf FindMaximum}.

\subsection*{Parameter Tables \& Fit Criteria Values} It should be noted that the lower-bound parameter, $k$, creates a discontinuity in the mixture {\it pdf} as constructed in this paper.  Hence, the numerical methods used to find the ML parameter estimates had great difficulty converging to a robust result when $k$ was not specified.  To circumvent this problem, $k$ was estimated using manual successive runs of estimating the remaining parameters and then adjusting $k$.  Consequently, no standard error estimate for $k$ is provided and the reader should consider its value taken as a prior rather than estimated.  The ML estimates were calculated using numerical maximization routines in {\it Mathematica}.

\clearpage

\begin{table}
	\centering
		\resizebox{5in}{!}{
		\begin{tabular}{>{\small} l>{\small} c>{\small} c>{\small} c} \hline
			{\bf Year} & {\bf Exp / Pwr-Law} & {\bf lg-Normal / Pwr-Law} & {\bf Exp / lg-Normal / Pwr-Law} \\
			 & $A$, $\beta$, $k = 44\textrm{k}, \alpha$ & $A$, $\mu,\sigma$, $k = 46\textrm{k}, \alpha$ & $A$, $B$, $\beta$, $\mu,\sigma$, $k = 45\textrm{k}, \alpha$\\[2pt] \hline
			\multicolumn{4}{c}{{\bf All Respondents}} \\[4pt]
			1998 & $0.944, 0.0000298, 1.033$ & $0.925, 9.98, 1.119, 4.122$ & $0.590, 0.080, 0.0000379, 10.35, 0.825, 1.266$ \\
			 & $(10^{-4}), (10^{-9}),(10^{-3})$ & $(10^{-4}), (10^{-4}), (10^{-4}), (10^{-3})$ & $(10^{-4}), (10^{-4}), (10^{-8}), (10^{-3}), (10^{-4}), (10^{-3})$ \\[2pt]
			2003 & $0.950, 0.0000257, 1.000$ & $0.910, 10.08, 1.120, 2.760$ & $0.626, 0.049, 0.0000293, 10.43, 0.910, 1.112$ \\
			 & $(10 ^{-4}), (10 ^{-9}), (10 ^{-3})$ & $(10 ^{-4}), (10 ^{-4}), (10 ^{-4}), (10 ^{-3})$ & $(10^{-4}), (10^{-4}), (10^{-8}), (10^{-3}), (10^{-4}), (10^{-3})$ \\[2pt]
			2008 & $0.945, 0.0000221, 0.882$ & $0.883, 10.18, 1.147, 1.704$ & $0.666, 0.028, 0.0000242, 10.60, 1.046, 0.875$ \\
			 & $(10 ^{-4}), (10 ^{-9}), (10 ^{-3})$ & $(10 ^{-4}), (10 ^{-4}), (10 ^{-4}), (10 ^{-3})$ & $(10^{-4}), (10^{-4}), (10^{-8}), (10^{-3}), (10^{-4}), (10^{-3})$ \\[4pt] \hline
			\multicolumn{4}{c}{{\bf Singles}} \\[4pt]
			1998 & $0.984, 0.0000488, 0.944$ & $0.985, 9.49, 1.067, 25.1$ & $0.704, 0.025, 0.0000686, 10.25, 0.615, 1.199$ \\
			 & $(10^{-5}), (10^{-8}),(10^{-3})$ & $(10^{-4}), (10^{-4}), (10^{-4}), (10^{-1})$ & $(10^{-3}), (10^{-4}), (10^{-7}), (10^{-4}), (10^{-3}), (10^{-3})$ \\[2pt]
			2003 & $0.982, 0.0000429, 1.002$ & $0.964, 9.59, 1.072, 9.64$ & $0.789, 0.025, 0.0000521, 10.36, 0.650, 1.191$ \\
			 & $(10^{-5}), (10^{-8}),(10^{-3})$ & $(10^{-4}), (10^{-4}), (10^{-4}), (10^{-2})$ & $(10^{-3}), (10^{-4}), (10^{-7}), (10^{-3}), (10^{-3}), (10^{-3})$ \\[2pt]
			2008 & $0.968, 0.0000375, 1.117$ & $0.971, 9.71,1.116, 7.38$ & $0.759, 0.040, 0.0000480, 10.56, 0.632, 1.243$ \\
			 & $(10^{-4}), (10^{-8}),(10^{-3})$ & $(10^{-4}), (10^{-4}), (10^{-4}), (10^{-2})$ & $(10^{-3}), (10^{-4}), (10^{-7}), (10^{-3}), (10^{-3}), (10^{-3})$ \\[2pt] \hline
		\end{tabular}}
	\caption{The table lists the ML parameter estimates for the various pooled distributions fitted to the IRS income data.  Upper limits to the estimated order of magnitude of the standard errors based on the second derivative of the log-likelihood function are given in parentheses.}
	\label{tbl:ParamEst}
\end{table}

\begin{table}
	\centering
		\resizebox{4in}{!}{
		\begin{tabular}{>{\small} l>{\small} l>{\small} l>{\small} l} \hline
			{\bf Year} & {\bf Exp / Pwr-Law} & {\bf lg-Normal / Pwr-Law} & {\bf Exp / lg-Normal / Pwr-Law} \\[2pt] \hline
			\multicolumn{4}{c}{{\bf All Respondents}} \\[4pt]
			\multirow{4}{*}{1998} & $ID = 0.0078$ & $ID = 0.0127$ & $ID = 0.0018$ \\
			 & $D_{KL} = 0.0078$ & $D_{KL} = 0.0128$ & $D_{KL} = 0.0018$ \\ 
			 & $D_{KS} = 347$ & $D_{KS} = 317$ & $D_{KS} = 24.2$ \\
			 & $AIC/BIC = 6.91 \times 10^{8}$ & $AIC/BIC = 6.93 \times 10^{8}$ & $AIC/BIC = 6.90 \times 10^{8}$ \\[4pt]
			\multirow{4}{*}{2003} & $ID = 0.0079$ & $ID = 0.0132$ & $ID = 0.0022$ \\
			 & $D_{KL} = 0.0079$ & $D_{KL} = 0.0133$ & $D_{KL} = 0.0022$ \\ 
			 & $D_{KS} = 311$ & $D_{KS} = 310$ & $D_{KS} = 42.3$ \\
			 & $AIC/BIC = 6.29 \times 10^{8}$ & $AIC/BIC = 6.32 \times 10^{8}$ & $AIC/BIC = 6.28 \times 10^{8}$ \\[4pt]
			\multirow{4}{*}{2008} & $ID = 0.0079$ & $ID = 0.0138$ & $ID = 0.0026$ \\
			 & $D_{KL} = 0.0079$ & $D_{KL} = 0.0139$ & $D_{KL} = 0.0026$ \\ 
			 & $D_{KS} = 202$ & $D_{KS} = 312$ & $D_{KS} = 46.7$ \\
			 & $AIC/BIC = 6.01 \times 10^{8}$ & $AIC/BIC = 6.03 \times 10^{8}$ & $AIC/BIC = 5.99 \times 10^{8}$ \\[4pt] \hline
			\multicolumn{4}{c}{{\bf Singles}} \\[4pt]
			\multirow{4}{*}{1998} & $ID = 0.0040$ & $ID = 0.0212$ & $ID = 0.0014$ \\
			 & $D_{KL} = 0.0040$ & $D_{KL} = 0.0214$ & $D_{KL} = 0.0014$ \\ 
			 & $D_{KS} = 123$ & $D_{KS} = 321$ & $D_{KS} = 29.0$ \\
			 & $AIC/BIC = 2.425 \times 10^{8}$ & $AIC/BIC = 2.445 \times 10^{8}$ & $AIC/BIC = 2.422 \times 10^{8}$ \\[4pt]
			\multirow{4}{*}{2003} & $ID = 0.0038$ & $ID = 0.0191$ & $ID = 0.0022$ \\
			 & $D_{KL} = 0.0038$ & $D_{KL} = 0.0193$ & $D_{KL} = 0.0022$ \\ 
			 & $D_{KS} = 114$ & $D_{KS} = 290$ & $D_{KS} = 54.6$ \\
			 & $AIC/BIC = 2.546 \times 10^{8}$ & $AIC/BIC = 2.565 \times 10^{8}$ & $AIC/BIC = 2.544 \times 10^{8}$ \\[4pt]
			\multirow{4}{*}{2008} & $ID = 0.0030$ & $ID = 0.0179$ & $ID = 0.0017$ \\
			 & $D_{KL} = 0.0030$ & $D_{KL} = 0.0181$ & $D_{KL} = 0.0017$ \\ 
			 & $D_{KS} = 96.3$ & $D_{KS} = 300$ & $D_{KS} = 47.3$ \\
			 & $AIC/BIC = 2.965 \times 10^{8}$ & $AIC/BIC = 2.987 \times 10^{8}$ & $AIC/BIC = 2.964 \times 10^{8}$ \\[4pt] \hline
		\end{tabular}}
	\caption{Fit criteria evaluated for the given pooled distributions fit to the IRS income data.  Note that the value of the log-likelihood dominates the $AIC$ and $BIC$ so that to the level of precision listed, they appear identical.}
	\label{tbl:FitCrit}
\end{table}

\end{document}